# ON NONLINEAR APPROXIMATIONS TO COSMIC PROBLEMS WITH MIXED BOUNDARY CONDITIONS

P. J. MANCINELLI[1], A. YAHIL[1], G. GANON[2], A. DEKEL[2]
[1]*Astronomy Program, State University of New York, Stony Brook, NY 11794-2100, U.S.A.*
[2]*Racah Institute, Hebrew University, Jerusalem, ISRAEL*

**Abstract**

Nonlinear approximations to problems with mixed boundary conditions are useful for predicting large-scale streaming velocities from the density field, or vice-versa. We evaluate the schemes of Bernardeau [1], Gramann [3], and Nusser *et al.* [5], using smoothed density and velocity fields obtained from $N$-body simulations of a CDM universe. The approximation of Nusser *et al.* is overall the most accurate and robust. For Gaussian smoothing of 1000 km s$^{-1}$ the mean error in the approximated relative density perturbation, $\delta$, is smaller than 0.06, and the dispersion is 0.1. The *r.m.s.* error in the estimated velocity is smaller than 60 km s$^{-1}$, and the dispersion is 40 km s$^{-1}$. For smoothing of 500 km s$^{-1}$ these numbers increase by about a factor $\sim 2$ for $\delta < 4-5$, but deteriorate at higher densities. The other approximations are comparable to those of Nusser *et al.* for smoothing of 1000 km s$^{-1}$, but are much less successful for the smaller smoothing of 500 km s$^{-1}$.

## 1 Introduction

Comparisons of large-scale density and velocity fields require solutions to problems with mixed boundary conditions (MBC). Unlike an initial value problem, in which positions and velocities are given at some initial epoch, and the time evolution of the system can then be computed, an MBC problem provides some of the positions and velocities at one epoch and some at another epoch.

For example, the density field, i.e., the positions, may be given at the present epoch, and the assumption of a growing gravitational instability requires the peculiar velocities to vanish in the limit of high redshift. With these boundary conditions one then seeks to compute the velocities at the present epoch. Conversely, the velocities may be given at the present epoch, and with the same assumption of a growing gravitational instability one seeks to compute the present day density field. A more complex case is one in which the density at the present epoch is given in redshift space, a combination of positions and velocities.

Nonlinear MBC problems have multi-valued solutions. For example, consider the two-body problem of the Galaxy and M31. On the basis of its position alone we can not tell if M31 is on its initial expansion away from the Galaxy, infalling after first apgalacticon, or passing by on a subsequent orbit. All these are valid solutions, and further criteria are needed to identify the correct one. Orbit crossing makes this identification more difficult, and it becomes impossible after virialization, which erases the memory of the initial conditions. In practice, one is therefore often restricted to laminar flow in the quasi-linear regime, in which perturbations are no longer small, but in which there has been no orbit crossing, and there exists a smooth one-to-one correspondence between the final and initial positions.

The Kelvin circulation theorem guarantees that the flow, which was initially irrotational in the linear domain, continues to be so in the nonlinear regime, as long as there is no orbit crossing, and the flow remains laminar. The velocity field is therefore completely specified by its divergence or, alternatively, by a potential.

## 2 Nonlinear Approximations

Initial value problems are now routinely computed into the nonlinear regime using $N$-body codes. There are also various shortcut approximations, some of which have been presented in this conference. But all initial value schemes do poorly in matching mixed boundary conditions. We will therefore consider only nonlinear approximations which are specifically designed for MBC problems.

The solution to the linear MBC problem is well known and unique [6]:

$$\nabla \cdot \mathbf{v} = -f(\Omega, \lambda)\delta \quad , \tag{1}$$

where $\mathbf{v}$ is the peculiar velocity field, $\delta \equiv \delta\rho/\rho$ is the relative mass perturbation, and we measure distances by their Hubble velocities, i.e., set $H_0 = 1$. The factor $f$ is the logarithmic derivative of the linear growth function, $D$, with respect to the expansion scale of the universe, $R$

$$f(\Omega, \lambda) = \frac{d \ln D}{d \ln R} \approx \Omega^{0.6} + \frac{\lambda}{70}\left(1 + \frac{\Omega}{2}\right) \quad . \tag{2}$$

As can be seen from Eq. (2), it depends mainly on the cosmological density parameter $\Omega$ [6] and only very weakly on the cosmological constant $\lambda$ [4], so henceforth we ignore $\lambda$.

Peebles [7, 8] pointed out that the general, nonlinear MBC problem lends itself naturally to an application of Hamilton's action principle. In this method the orbit of each mass point is approximated as a sum of given functions of time, multiplied by coefficients which are determined by finding extrema of the action. Giavalisco *et al.* [2] have recently shown that the orbits can be usefully approximated as polynomials in the linear growth function. The MBC problem can be solved by the action method for any geometry, and to any desired accuracy, depending on the order of the polynomial. In fact, Giavalisco *et al.* obtained excellent fits to the nonlinear spherical case, the most difficult for this scheme, using second and third orders.

Three shortcut schemes to nonlinear MBC problems have been proposed over the last two years which make no assumption about geometrical symmetries [1, 3, 5]. They are all local, in the sense that they either approximate the relative density perturbation, $\delta$, as explicit functions of the deformation tensor, $\partial v_i/\partial x_j$ or, conversely, approximate $\nabla \cdot \mathbf{v}$ as an explicit function of gradients of the peculiar gravity, $\partial g_i/\partial x_j$. We shall henceforth refer to them as *local* approximations.

The Bernardeau approximation [1] was derived from perturbation theory in the limit of vanishing variance of the density field

$$\nabla \cdot \mathbf{v} = \frac{3}{2} f(\Omega)\left[1 - (1+\delta)^{2/3}\right] \quad . \tag{3}$$

This relation is invertible, so the velocity may be determined from the density, or vice versa.

Gramann [3] derived the second order Lagrangian perturbation approximation

$$\nabla \cdot \mathbf{v} = -f(\Omega)\left[\delta - \frac{16}{63\Omega^2} m_g\right] \quad , \tag{4}$$

where

$$m_g = \sum_{i<j} \left[ \frac{\partial g_i}{\partial x_i} \frac{\partial g_j}{\partial x_j} - \left( \frac{\partial g_i}{\partial x_j} \right)^2 \right] \quad . \tag{5}$$

Since $\mathbf{g}$ is an explicit function of $\delta$, Eq. (4) provides an approximation for $\mathbf{v}$, given $\delta$. The second-order inverse relation is

$$\delta = -\frac{\nabla \cdot \mathbf{v}}{f(\Omega)} + \frac{4}{7} \frac{m_v}{f(\Omega)^2} \quad , \tag{6}$$

where $m_v$ is analogous to $m_g$, with $\mathbf{g}$ replaced by $\mathbf{v}$.

Nusser *et al.* [5] generalized the Zel'dovich [9] approximation to Eulerian coordinates, and expressed the continuity equation as

$$\delta = \left\| \delta_{ij} - f(\Omega)^{-1} \frac{\partial v_i}{\partial x_j} \right\| - 1 \quad , \tag{7}$$

where the double vertical lines denote the Jacobian determinant. This relation provides the density as a function of velocity, but is difficult to invert. Instead, the authors used a phenomenological formula:

$$\nabla \cdot \mathbf{v} = -\frac{f(\Omega)\delta}{1 + 0.18\delta} \quad . \tag{8}$$

## 3 Evaluation Using N-Body Simulations

We tested the local nonlinear approximations described in §2 using three independent $N$-body simulations of a CDM universe with $\Omega = 1$ and $h = 0.5$ ($H_0 = 100h$ km s$^{-1}$ Mpc$^{-1}$), which we integrated from the linear regime at high redshift to the present epoch. The simulations, which were identical except for the choice of the seed of the random number generator used to create the initial conditions, were run on a $128^3$ grid with comoving spacing of 200 km s$^{-1}$. The initial perturbations were normalized in the standard way to unit variance in a sphere of radius 800 km s$^{-1}$, if extrapolated linearly to the present epoch. They were then integrated forward using a particle-mesh (PM) code kindly provided by E. Bertschinger. The resultant, fully nonlinear, density and velocity fields at the present epoch were computed at the grid points using a cloud-in-cell method, followed by Gaussian smoothing; we refer to them as the "exact" fields.

The assumption of potential (irrotational) flow is central to all the local approximations considered, so we first check how well the exact velocity field satisfies this condition. On small scales there is, of course, orbit crossing. Sufficiently smoothed flow, however, is expected to be laminar and hence irrotational. Fig. 1 shows the frequency distribution of $|\nabla \times \mathbf{v}|$ for Gaussian smoothing radii of 500 km s$^{-1}$ and 1000 km s$^{-1}$. Recall that we measure distance in km s$^{-1}$, so $\nabla \times \mathbf{v}$ is a dimensionless quantity, which Fig. 1 shows to be on the order of 0.01. This is to be compared with $\nabla \cdot \mathbf{v} \sim \delta \sim 1$. We conclude that deviations from potential flow are at the 1% level.

In Figs. 2 and 3 we compare the approximate fields that were obtained from the exact ones using the various approximations of §2. The approximations for the density field, given a velocity field, are evaluated in the lower panels, in which we plot the differences between the approximate and exact densities, $\Delta\delta$. The points are the *mean* differences for each of the three simulations and show the systematic error of each approximation as a function of the (exact) density. The error bars in the figures are a measure of the dispersion in the approximations of *individual* data (grid) points. They were measured by taking the *r.m.s.* cross differences of $\Delta\delta$ between points in independent simulations. Note that grid points in the same simulation may not be used for this purpose, because they may fall within each other's smoothing distance, causing an underestimate of the dispersion. For the same reason, the standard deviations of the means are not equal to the standard deviations divided by $\sqrt{N}$, nor are the means of different bins independent. It is therefore better to estimate the uncertainties in the means from the scatter between different simulations, and we have chosen simply to plot the means of all three simulations.

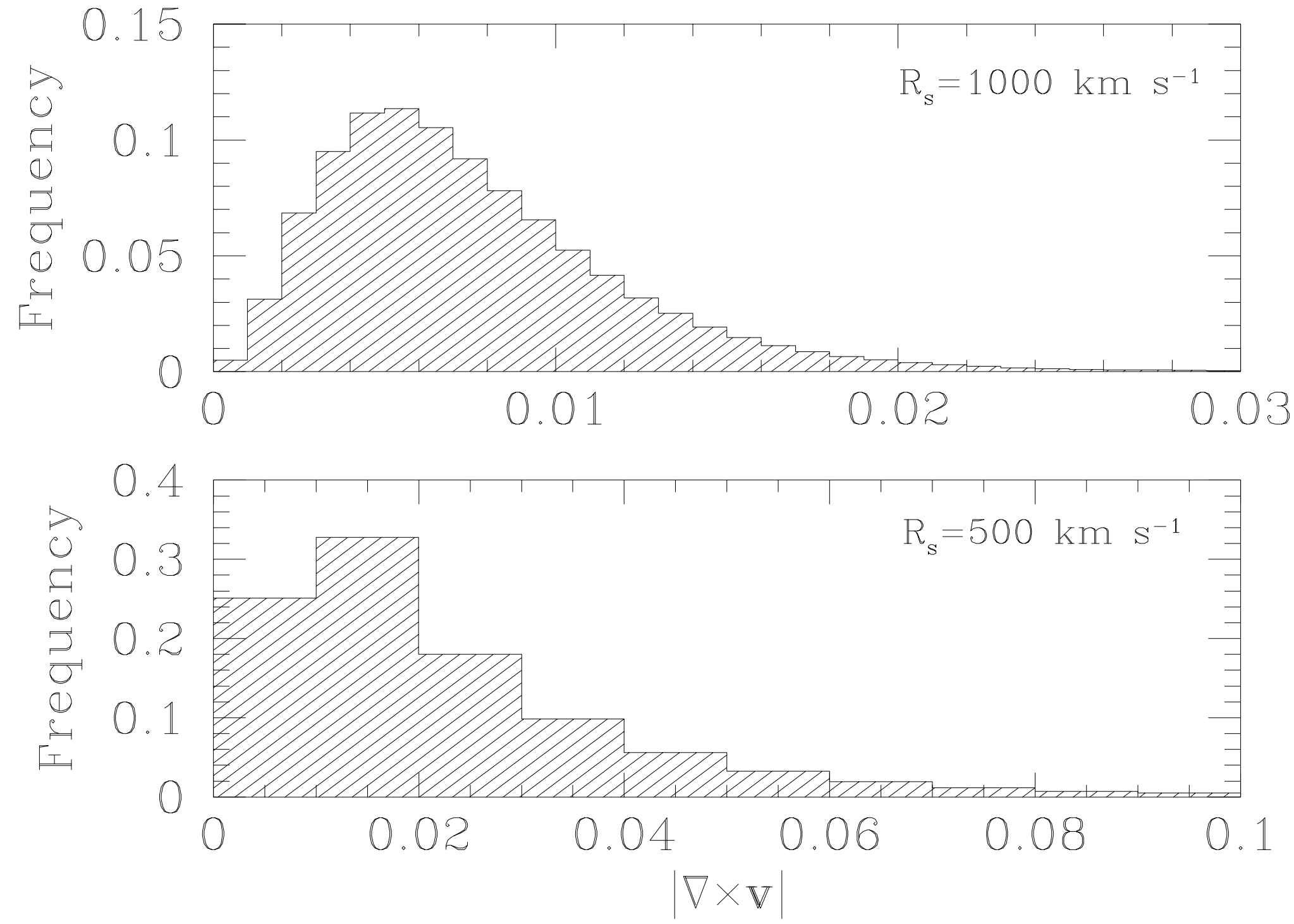

Figure 1: Frequency distribution of $|\nabla \times \mathbf{v}|$ in $N$-body simulations of a CDM universe, for Gaussian smoothing radii of 500 km s$^{-1}$ and 1000 km s$^{-1}$. The deviations from potential flow are $\sim 1\%$.

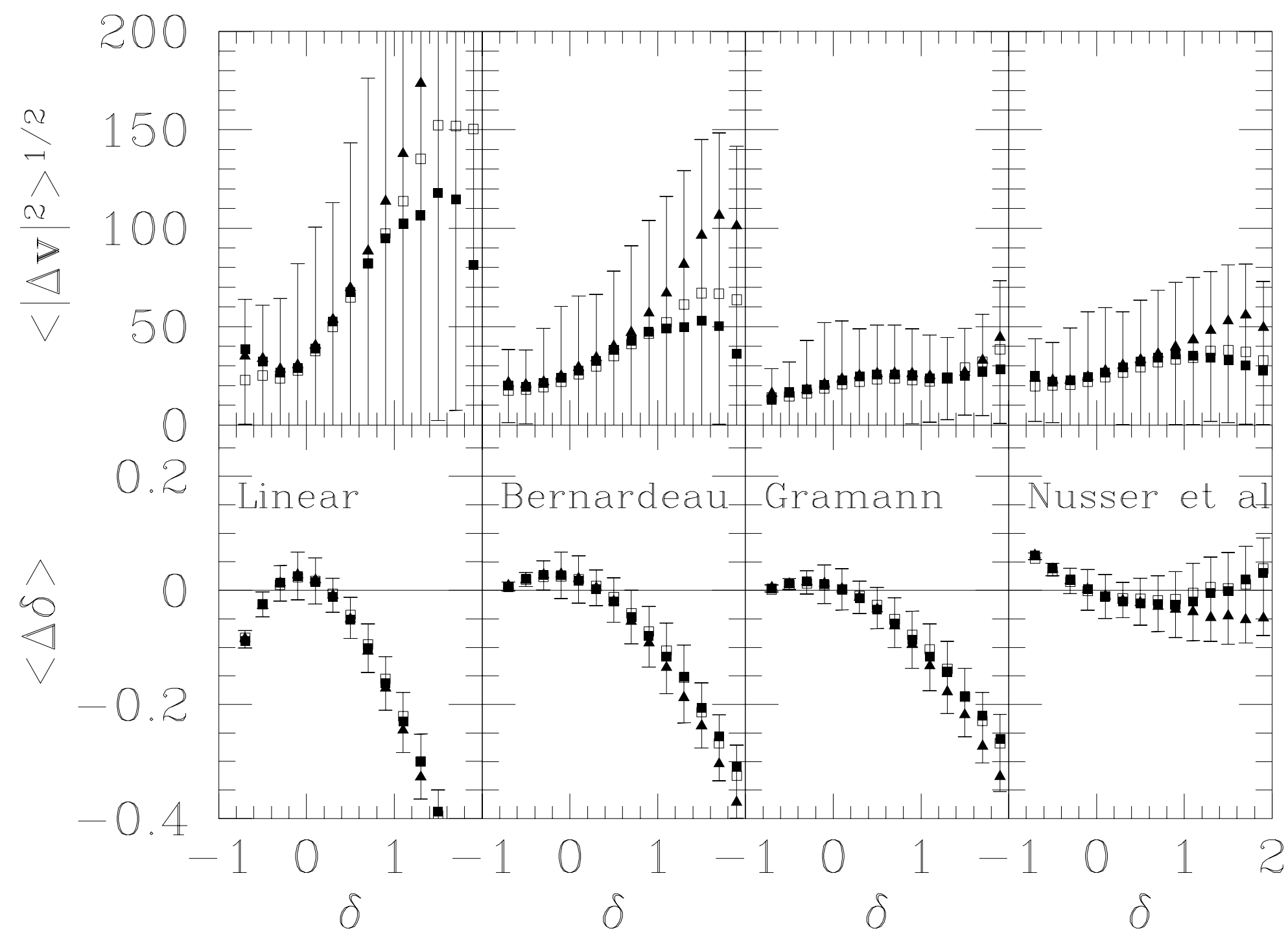

Figure 2: Differences between the nonlinear approximations of §2 and the "exact" fields obtained from $N$-body simulations of a CDM universe, Gaussian smoothed with a radius of 1000 km s$^{-1}$. The points are the *mean* values obtained in each density bin in three identical simulations, differing only in the seed of the random number generator which created the initial conditions. The error bars are a measure of the dispersion in the approximations of *individual* data (grid) points. See the text for details.

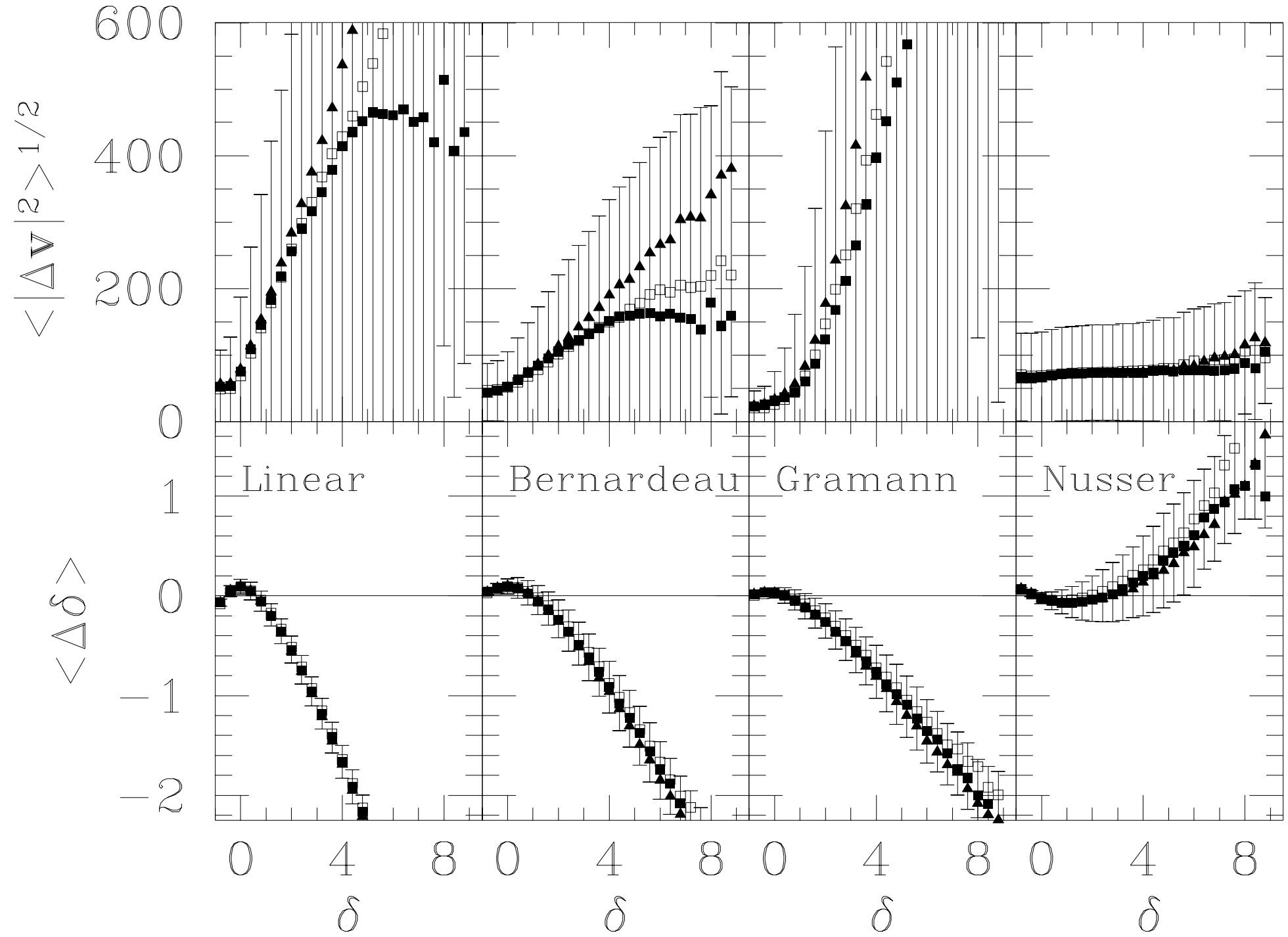

Figure 3: Same as Fig. 2 with Gaussian smoothing of 500 km s$^{-1}$.

The upper panels show the analogous comparison for velocity fields approximated from densities. The quantity for comparison here is $|\Delta\mathbf{v}|^2$, except that we always plot its square root, in order to express it in km s$^{-1}$.

Fig. 2 shows that, with a smoothing of 1000 km s$^{-1}$, $\delta < 2$, which may be compared with the turn-around density of a top-hat perturbation, $\delta = 4.6$. While already nonlinear, these perturbations are still mild, and all the nonlinear approximations are fairly accurate. The approximation of Nusser *et al.* [5] does best, with a mean deviation $|\langle\Delta\delta\rangle| < 0.06$ for density approximations, and $\langle|\Delta\mathbf{v}|^2\rangle^{1/2} <$ 60 km s$^{-1}$ for velocity approximations. The dispersions for individual grid points for these variables are smaller than 0.1 and 40 km s$^{-1}$, respectively. Note, however, that the density approximation of Gramann [3] is better than that of Nusser *et al.* for negative $\delta$, and her velocity approximation is slightly better for all $\delta$.

The situation changes for the smaller smoothing of 500 km s$^{-1}$, Fig. 3. Peak densities now reach $\delta \sim 10$. The density approximation of Nusser *et al.* continues to be very good up to $\delta \sim 4-5$, around the top-hat turn-around density, and then begins to break down. The other approximations break down much earlier. The velocity approximation of Nusser *et al.* is remarkably robust, with $\langle|\Delta\mathbf{v}|^2\rangle^{1/2} < 120$ km s$^{-1}$ and a dispersion of 80 km s$^{-1}$. By comparison, all the other approximations fare much worse, with that of Gramann not much better than linear.

We conclude that the approximations of Nusser *et al.* [5] are the most robust and accurate, with the density approximation beginning to break down around $\delta \sim 4-5$, but the velocity approximation holding to $\delta \sim 10$.

**Acknowledgements.** We thank E. Bertschinger for providing his $N$-body code.

## References

[1] Bernardeau, F. 1992, *Astrophys. J. Lett.*, 390, L61

[2] Giavalisco, M., Mancinelli, B., Mancinelli, P. J., & Yahil, A. 1993, *Astrophys. J.*, 411, 9

[3] Gramann, M. 1993, *Astrophys. J. Lett.*, 405, L47

[4] Lahav, O., Lilje, P. B., Primack, J. R., & Rees, M. J. 1991, *Mon. Not. R. Astr. Soc.*, 251, 128

[5] Nusser, A., Dekel, A., Bertschinger, E., & Blumenthal, G. R. 1991, *Astrophys. J.*, 379, 6

[6] Peebles, P. J. E. 1976, *Astrophys. J.*, 205, 318

[7] ———. 1989, *Astrophys. J. Lett.*, 344, L53

[8] ———. 1990, *Astrophys. J.*, 362, 1

[9] Zel'dovich, Ya. B. 1970, *Astrofizika*, 6, 319; *Astr. Astrophys.*, 5, 84